\documentclass{aa}  
\usepackage{natbib} 
\usepackage{graphicx}
\usepackage{txfonts}
\usepackage{lipsum}
\usepackage{subcaption}         % necessary for continued figures, example in section 3
\usepackage{lscape}             % to rotate a single page table, example in appendix.
\usepackage{placeins}           % useful with \FloatBarrier, to keep 
\begin{document}

%%%%%%%%%%%%%%%%%%%%%%%%%%%%%%%%%%%%%%%%
% if you use custom commands in your title,
% ensure to check your title when submitting!
%%%%%%%%%%%%%%%%%%%%%%%%%%%%%%%%%%%%%%%%
   \title{A Gravitationally Lensed Low-Luminosity AGN From The Cosmic Noon}

   \subtitle{}

%%%%%%%%%%%%%%%%%%%%%%%%%%%%%%%%%%%%%%%%
% Please separate each author with the \and command
%
% Use the \corrauth to provide the corresponding
% author address. It will be automatically inserted as 
% footnote in the PDF output.
%
% Please DO NOT include ORCIDs next to author names.
% Instead, please provide an active address for each coauthor:
% it will be automatically extracted by EDPS editorial system, 
% and co-authors will be be able to authenticate their ORCID.
%
% Only authenticated ORCIDs will be taken into account.
% ORCIDs included here will be removed.
%%%%%%%%%%%%%%%%%%%%%%%%%%%%%%%%%%%%%%%%

   \author{M. Mi\'ci\'c\inst{1}\corrauth{micic@ou.edu}        % use \corrauth for the corresponding author
        \and O. Holmes\inst{2}\email{}
        \and X. Dai\inst{1}\email{}
        \and K. Hainline\inst{3}\email{}
        \and H. Xue\inst{1}\email{}
        \and Z. Gao\inst{1}\email{}
        }

   \institute{University of Oklahoma, Norman, OK, USA
   \and Purdue University, West Lafayette, IN, USA
   \and University of Arizona, Tucson, AZ, USA
   }

   \date{Received September 30, 20XX}

% \abstract{}{}{}{}{}
% 5 {} token are mandatory
 
 \abstract
   {Strong gravitational lensing provides a unique tool of studying intrinsically faint high-redshift active galactic nuclei (AGN). In particular, low-luminosity AGN (LLAGN) can probe black-hole growth in low-mass galaxies that current X-ray surveys cannot otherwise access.}
   {We investigate the nature and physical properties of a faint X-ray source triply imaged by the galaxy cluster SDSS J2243$-$0935, and determine whether the source hosts an AGN.} 
   {We combine archival Hubble Space Telescope imaging and Chandra X-ray observations with the published strong-lensing models of SDSS J2243$-$0935. We constrain the source redshift geometrically, derive magnification maps, and infer the physical properties of the host galaxy.}
   {The geometric lensing analysis gives a source redshift of $z=2.14^{+0.12}_{-0.17}$, consistent with the photometric redshift estimate, and a total magnification of $\mu=78^{+7}_{-6}$. The source is triply lensed and detected only in the HST F160W band, with an intrinsic magnitude of $m_{\rm AB}=28.2\pm0.2$. The rest-frame optical emission implies a stellar mass of $8.6\lesssim\log(M_\star/M_\odot)\lesssim9.4$. The unobscured star formation rate is constrained to $\mathrm{SFR}_{\rm UV}<0.009\,M_\odot\,\mathrm{yr}^{-1}$, although substantial dust obscuration remains possible. Chandra detects X-ray point-source counterparts to all three HST-detected lensed images. The combined X-ray spectrum contains $100\pm11$ net counts and is consistent with a moderately absorbed power law, with $\Gamma=1.48^{+0.28}_{-0.25}$ and $N_{\rm H}=(5.5^{+5.2}_{-3.6})\times10^{22}\,\mathrm{cm}^{-2}$. After correcting for gravitational magnification, we obtain an intrinsic rest-frame $2$--$10$ keV luminosity of $L_{\rm X}=1.2^{+0.3}_{-0.2}\times10^{42}\,\mathrm{erg\,s^{-1}}$.}
   {The X-ray emission is likely dominated by an AGN hosted by a low-mass galaxy at $z\simeq2.14$, making this system a strongly lensed LLAGN at cosmic noon. Its large magnification provides an effective X-ray exposure equivalent to approximately 9~Ms, demonstrating the potential of strong lensing to reveal intrinsically faint black-hole accretion in distant low-mass galaxies.}


   \maketitle

\nolinenumbers
%%%%%%%%%%%%%%%%%%%%%%%%%%%%%%%%%%%%%%%%%%%%%%%%%%%%%%%%%%%%%%
\section{Introduction}
Strong gravitational lensing has become an indispensable tool for probing the distant universe, enabling studies of sources that would otherwise be too faint to observe directly. In the X-ray regime, however, the population of known multiply imaged sources is dominated by quasars \citep[e.g.,][]{2000ApJ...531...81C,2004ApJ...605...45D,2012ApJ...755...24C,2017ApJ...836..206G}, which already have high intrinsic X-ray luminosities. Despite hundreds of galaxy--galaxy strong lenses discovered in large-scale optical surveys \citep{2006ApJ...638..703B,2012ApJ...744...41B}, multiply imaged intrinsically faint X-ray sources remain exceedingly rare. Only a handful of systems associated with populations such as normal star-forming galaxies \citep{2020NatAs...4..159B} or low-luminosity active galactic nuclei (LLAGN; $L_{\rm X}<10^{42}$ erg s$^{-1}$) have been identified.

Detections of high-$z$ LLAGN are significant for studies of early black hole growth and the coevolution between galaxies and their central engines \citep[e.g.,][]{10.1093/mnras/stad1877,2018MNRAS.478.2576M}. Most AGN detected at high redshift are luminous systems powered by rapidly accreting supermassive black holes, and therefore preferentially trace black holes that have already experienced substantial growth. LLAGN instead probe the faint end of the accreting population, potentially revealing lower-mass black holes and systems undergoing more modest episodes of growth \citep{2018ApJ...863....1C}. Although several high-$z$ LLAGN candidates have recently been identified with JWST \citep{2023ApJ...942L..17O, 2023ApJ...954L...4K}, establishing their X-ray properties remains challenging because of their intrinsic faintness and limited sensitivity of current facilities. Even the deepest X-ray observations to date, such as the \textit{Chandra} Deep Field South, have not yielded reliable LLAGN detections at $z>2$ \citep{2017ApJS..232....8L}.

Gravitational magnification offers a way to overcome this obstacle. In this work, we report the X-ray detection of a triply imaged LLAGN at $z=2.14$, strongly lensed by the foreground galaxy cluster SDSS J2243$-$0935 at $z=0.447$. With a total magnification of $\mu=78\pm6$, the system provides a rare opportunity to probe intrinsically faint black hole accretion at cosmic noon through X-ray emission that would otherwise lie below the detection threshold of \textit{Chandra}. We combine the lensing reconstruction with optical/near-infrared imaging and X-ray observations to characterize the intrinsic properties of the source and investigate the nature of its nuclear activity.

%%%%%%%%%%%%%%%%%%%%%%%%%%%%%%%%%%%%%%%%%%%%%%%%%%%%%%%%%%%%%%
\section{Data}
\begin{figure*}
        \centering
        \includegraphics[width=\hsize]{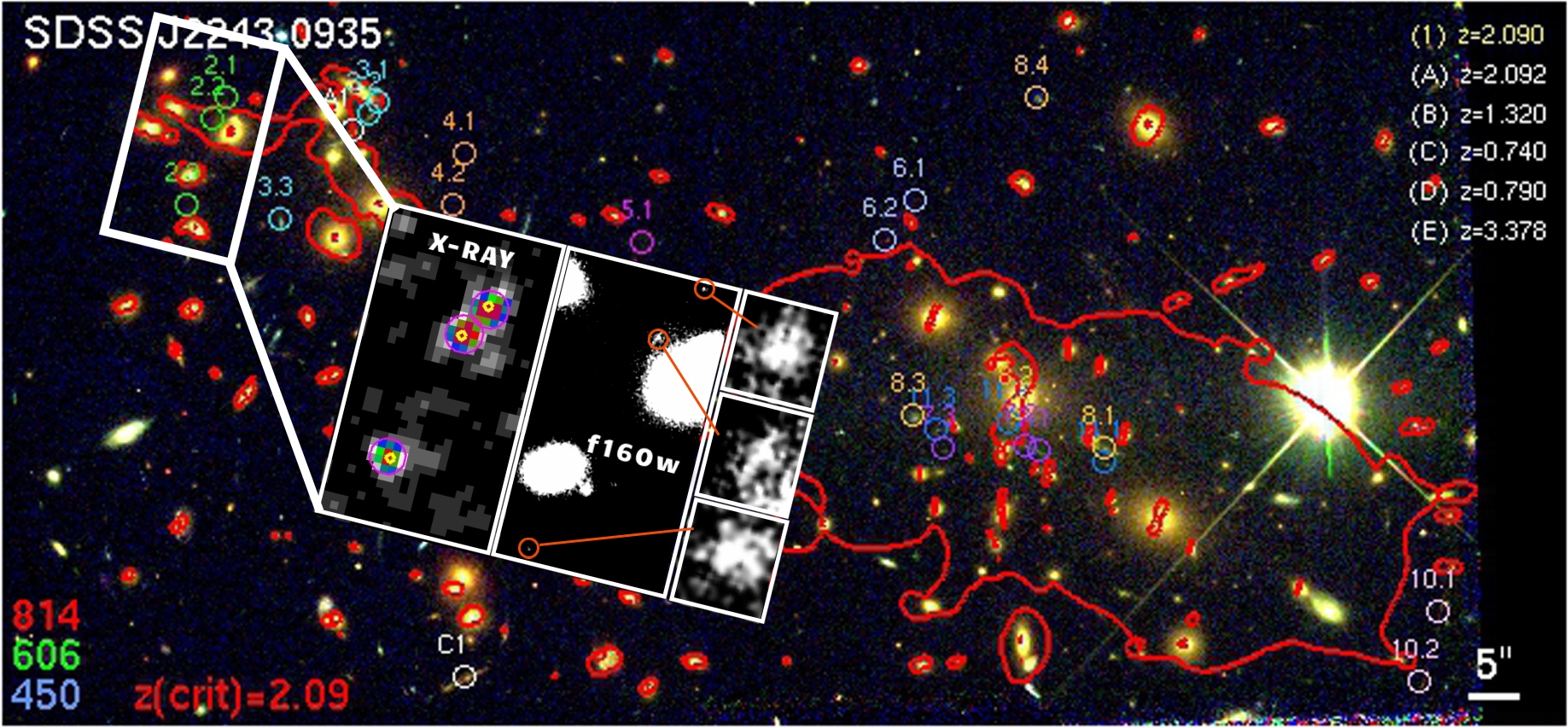}
        {\caption{The background image is an HST image of the cluster SDSS J2243$-$0935 from \citet{2020ApJS..247...12S}. The image shows eleven multiply lensed images labeled with colored circles and numbers 1.$i$ through 11.$i$. The white rectangle in the upper left corner shows the region of interest and the triply lensed source labeled with green circles and numbers 2.1, 2.2, and 2.3. The inset image is a zoomed-in region of interest shown in X-ray and HST F160W filters. The left panel shows the $Chandra$ X-ray image and three X-ray sources. Purple circles show 1.$\farcs$2 radius source extraction regions, while yellow symbols show HST-detected triply lensed images corresponding to 2.1, 2.2, and 2.3 from the background image. The middle panel shows the same region but in the HST F160W filter. The circled sources are triply lensed images. The right panel shows each image zoomed in.}
        \label{fig:optxray}}
\end{figure*}
The multiply imaged system studied in this work was originally identified using the Hubble Space Telescope (HST) in the SDSS Giant Arc Survey (SGAS), a systematic search for strong gravitational lensing by galaxy clusters in the Sloan Digital Sky Survey \citep[][hereafter S20]{2020ApJS..247...12S}. SGAS targets massive clusters whose gravitational potentials produce highly magnified arcs and multiple images of background galaxies, enabling studies of faint, high-redshift sources.

S20 used HST observations in six bands, including F390W, F450W, F606W, F814W, F105W, and F160W, to construct strong-lensing models of the galaxy cluster SDSS J2243\--0935, resulting in the detection of eleven multiply lensed background sources. We use the publicly available reduced HST data and lens output models provided by S20\footnote[1]{https://archive.stsci.edu/hlsp/sgas} to conduct the redshift analysis, produce magnification maps, and reconstruct the properties of background sources. 

We additionally analyze archival \textit{Chandra X-ray Observatory} observations covering the cluster field. SDSS J2243$-$0935 was observed three times with ACIS-I, for a combined exposure of 121 ks (ObsID 3260, PI: Van Speybroeck; ObsIDs 19749 and 20129, PI: Garmire). We use these observations to search for X-ray counterparts to the multiply imaged background sources and, as described below, identify X-ray emission spatially coincident with all three images of one lensed source. The HST color image of the cluster SDSS J2243\--0935, together with all eleven multiply lensed sources, and inset X-ray and HST F160W images of the region of interest are shown in Figure \ref{fig:optxray}.
\begin{figure*}
        \centering
        \includegraphics[width=\hsize]{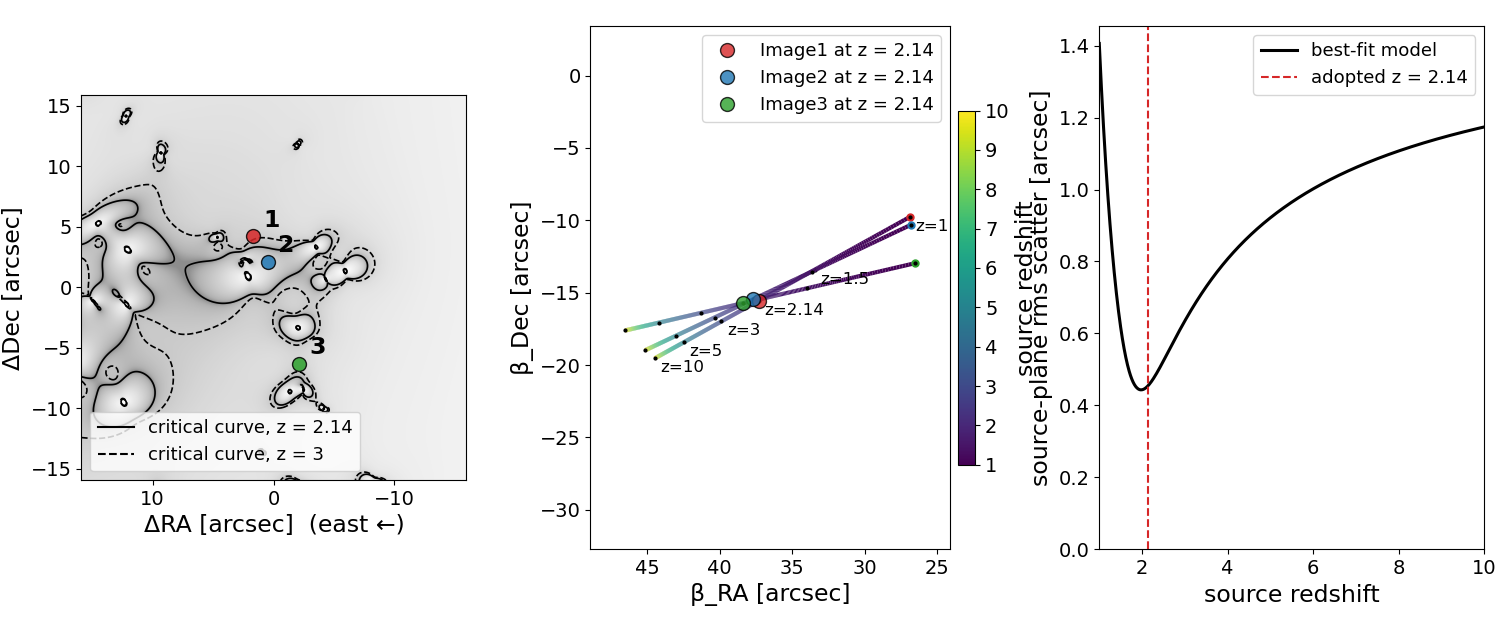}
        {\caption{Geometric redshift of the triply imaged background source, derived from the best-fit lens model of S20. \textit{Left}: Image plane, showing the positions of images 1, 2, and 3 relative to their centroid (east to the left, north up). The grayscale shows the magnification at $z = 2.14$, and the solid and dashed curves show the critical curves at $z = 2.14$ and $z = 3$, respectively. \textit{Middle}: Source plane. Each track shows the position to which one image maps back through the lens equation as the assumed source redshift increases from $z$ = 1 to 10, color-coded by redshift, with selected redshifts labeled. Large circles mark the back-projected positions at the adopted redshift, $z = 2.14$. At the true source redshift, all three images map back to the same position. \textit{Right}: rms scatter of the three back-projected positions as a function of source redshift for the best-fit model. The dashed red line marks the adopted redshift, $z = 2.14^{+0.12}_{-0.17}$, the median across all model realizations.}
        \label{fig3}}
\end{figure*}
\section{Results} \label{sec:3}
\subsection{Source redshift}
\label{sec:redshift}
S20 reports a photometric redshift of $z_{\rm phot}=3$ for the triply imaged source with X-ray counterparts. However, the source is robustly detected in only one of the available \textit{HST} bands, while the remaining bands provide weak detections or upper limits. Moreover, the reported photometric redshift is given without an associated uncertainty. For these reasons, we treat $z_{\rm phot}=3$ with caution and perform an independent analysis to constrain the redshift.

The available \textit{HST} photometry is characterized by a significant flux increase between F105W and F160W, while the source remains undetected or weakly detected at shorter wavelengths. Such a spectral break can potentially be interpreted as either a high-redshift ($z > 10$) Lyman-$\alpha$ break or a lower-redshift Balmer break, motivating us to consider both possibilities. If the Lyman-$\alpha$ break were the case, and we assume the F160W flux is tracing the rest-frame UV, the UV magnitude of the source would be M$_{\mathrm{UV}} \approx -19$, very similar to what is observed for galaxies at this redshift \citep{2026ApJ..1004..161H}. However, given the uncertainties on the fluxes, there is the possibility that the source may lie at a redshift where the observed break between the F105W and F160W filters is caused by a Balmer break in the host galaxy, with some level of dust obscuration or an older stellar population. To explore this conclusion, we matched the lensed sources to the colors and magnitudes for those observed in the JWST Advanced Deep Extragalactic Survey (JADES) DR5 photometric dataset \citep{2026arXiv260115956R} in GOODS-S. We explored a sample of over 500 sources with similar F105W -- F160W colors, F160W AB magnitudes, and F606W and F814W fluxes below the derived upper limits, finding that while the objects best matched included a sample of sources at $z > 9$, the bulk of the sources with similar colors were at $z \sim 2 - 3$ ($z_{\mathrm{med}} = 2.3$).

We additionally constrained the redshift geometrically using the strong-lensing model of S20. The model provides convergence maps normalized to $D_{\rm ls}/D_{\rm s}=1$, together with a set of MCMC realizations sampling the posterior uncertainty of the lens model. For each realization, we computed the deflection $\boldsymbol{\alpha}$ at each of the three observed image positions $\boldsymbol{\theta}$ from the convergence map, and back-projected the images to the source plane through the lens equation, $\boldsymbol{\beta} = \boldsymbol{\theta} - (D_{\rm ls}/D_{\rm s})\,\boldsymbol{\alpha}(\boldsymbol{\theta})$.
We surveyed the grid of redshifts spanning $1\leq z\leq10$ in steps of $\Delta z=0.01$. At each redshift, we calculated the rms separation of the three back-projected positions from their mean source-plane position and adopted the redshift that minimized this scatter. Repeating the procedure for all realizations yields a geometric redshift of $z_{\rm geom}=2.14^{+0.12}_{-0.17}$, which we adopt throughout the remainder of this work. The geometric-redshift determination is summarized in Figure~\ref{fig3}.

\subsection{Magnification maps}
\label{sec:magnification}
\begin{figure*}
        \centering
        \includegraphics[width=\hsize]{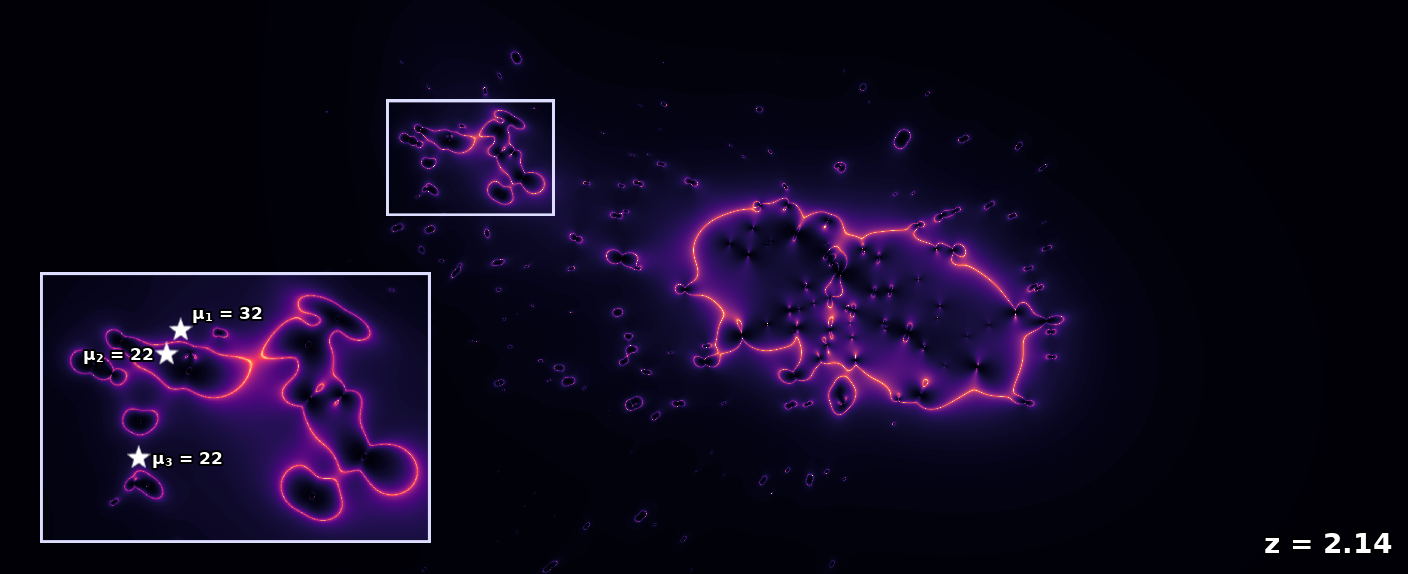}
        {\caption{Magnification map. The inset figure shows a zoomed-in region of interest, with white stars labeling locations of a triply lensed source.}
        \label{fig:magnmap}}
\end{figure*}

We combined the $\kappa$ and $\gamma$ maps with publicly available magnification tools\footnote[2]{https://archive.stsci.edu/prepds/frontier/lensmodels/} to construct the magnification map of the galaxy cluster SDSS J2243$-$0935. The resulting magnification map for a source at $z=2.14$ is shown in Figure \ref{fig:magnmap}. This yields $\mu_1 = 32.3^{+5.6}_{-4.5}$, $\mu_2 = 22.2^{+3.5}_{-2.6}$, and $\mu_3 = 22.0^{+6.0}_{-3.5}$. As an independent consistency check, the magnification-corrected F160W fluxes of the three images are consistent with one another within $1\sigma$.
\subsection{Photometry and intrinsic fluxes}
\label{sec:fluxes}

We measured fluxes in F105W, F160W, F606W, F450W, and F814W for images~1 and~3; image~2 was excluded because it is embedded in the light of a foreground galaxy. For each image, the local background was measured in an identical aperture on a nearby source-free region and subtracted, and count rates were converted to flux densities using the \texttt{PHOTFLAM} keyword. Because the drizzled images have correlated noise, we estimated the photometric uncertainties empirically from the difference between the two background measurements.

We combined the two images to estimate the intrinsic source flux as:
\begin{equation}
F_{\rm int} = \frac{F_1+F_3}{\mu_1+\mu_3},
\end{equation}
where $F_i$ and $\mu_i$ are the observed flux and magnification of each image, and $\mu_1 + \mu_3 = 54^{+9}_{-5}$. Uncertainties were propagated with Monte Carlo realizations sampling both the photometric errors and the lens-model magnifications. The source is detected only in F160W (S/N\,$\approx 7$), with an intrinsic flux density of $(2.38^{+0.49}_{-0.42}) \times 10^{-21}$\,erg\,s$^{-1}$\,cm$^{-2}$\,\AA$^{-1}$ ($m_{\rm AB} = 28.2 \pm 0.2$). For the other bands, we report $3\sigma$ upper limits. The summary of this analysis is given in Table \ref{tab:photometry}.
\begin{table*}
\centering
\caption{Photometry of the lensed source, with images~1 and~3 combined. Observed fluxes are the sum of the two images, $F_1 + F_3$; delensed fluxes are $(F_1 + F_3)/(\mu_1 + \mu_3)$, with uncertainties from a Monte Carlo over photometric errors and lens-model realizations at $z = 2.14$. Fluxes are in units of erg\,s$^{-1}$\,cm$^{-2}$\,\AA$^{-1}$. AB magnitudes use the pivot wavelength of each filter. For bands with S/N\,$<3$, magnitudes are $3\sigma$ limits; delensed limits use the $1\sigma$ lower bound on the total magnification ($\mu_1 + \mu_3 = 49$). Image~2 is excluded because it is blended with a foreground galaxy. All uncertainties are $1\sigma$.}
\label{tab:photometry}
\begin{tabular}{llcccccc}
\hline
Filter & Instrument & $\lambda_{\rm pivot}$ & $F_{\rm obs}$ & S/N & $m_{\rm obs}$ & $F_{\rm int}$ & $m_{\rm int}$ \\
 & & (\AA) & ($10^{-20}$) & & (AB) & ($10^{-22}$) & (AB) \\
\hline
F450W & WFPC2      & 4557  & $3.8 \pm 2.0$  & 1.9  & $>27.33$         & $6.9^{+3.9}_{-3.7}$  & $>31.56$ \\
F606W & WFC3/UVIS  & 5889  & $2.0 \pm 2.9$  & 0.7  & $>26.40$         & $3.7^{+5.3}_{-5.2}$  & $>30.63$ \\
F814W & WFPC2      & 7996  & $-0.3 \pm 2.1$ & $-0.1$ & $>26.08$       & $-0.5 \pm 3.8$       & $>30.30$ \\
F105W & WFC3/IR    & 10552 & $1.9 \pm 2.9$  & 0.7  & $>25.13$         & $3.4^{+5.3}_{-5.2}$  & $>29.36$ \\
F160W & WFC3/IR    & 15369 & $13.1 \pm 1.9$ & 6.9  & $23.87 \pm 0.16$ & $23.8^{+4.8}_{-4.3}$ & $28.22^{+0.21}_{-0.20}$ \\
\hline
\end{tabular}
\end{table*}
\subsection{X-ray analysis}
\label{sec:xray}

For the X-ray data analysis, we used \textsc{CIAO 4.17}, an interactive software package for observation analysis developed by the Chandra X-ray Center \citep{2006SPIE.6270E..1VF}. We used \textsc{CALDB 4.12.2}, the calibration database that stores and provides access to all calibration files required for standard processing and analysis. We reprocessed observations using the CIAO script \textsc{chandra\_repro}, merged all observations, and created point-spread function (PSF) maps using the \textsc{merge\_obs} script. The 90\% exposure-time-weighted PSF radius is $\sim$1\farcs2.

We detect three statistically significant X-ray point sources, spatially coincident with the three HST images of the lensed galaxy (Fig.~\ref{fig:optxray}). Source counts were extracted from circular apertures of radius $\sim$1\farcs2, and background from a source-free annulus with ten times the source-region area. The three images contain $100 \pm 11$ net counts in total. Accounting for the total magnification, the effective exposure time is $\sim$9\,Ms, exceeding that of the deepest \textit{Chandra} surveys.

We created response files for each observation with \textsc{specextract}, and combined the spectra with \textsc{combine\_spectra}. We fitted the 0.5--7\,keV spectrum in \textsc{Sherpa} \citep{2001SPIE.4477...76F} using the W-statistic, with the spectrum grouped to at least one count per bin. The model consisted of a power law absorbed by a fixed Galactic component, and an intrinsic component at $z = 2.14$ with free column density. A fit without intrinsic absorption yields a flat spectrum, $\Gamma = 1.19 \pm 0.18$. Because $\Gamma$ and the intrinsic $N_{\rm H}$ are degenerate at this signal-to-noise, we marginalized over both using MCMC analysis. This gives $\Gamma = 1.48^{+0.28}_{-0.25}$ and $N_{\rm H} = 5.5^{+5.2}_{-3.6} \times 10^{22}$\,cm$^{-2}$.

At $z = 2.14$, the rest-frame 2--10\,keV band corresponds to observed 0.64--3.19\,keV. Integrating the unabsorbed power law over this band gives $F_X = 2.78^{+0.70}_{-0.47} \times 10^{-15}$\,erg\,s$^{-1}$\,cm$^{-2}$. The intrinsic luminosity, $L_X = 4\pi d_L^2 F_X / \mu_{\rm tot}$, with uncertainties propagated from both the spectral fit and the magnification, is $L_X = 1.2^{+0.3}_{-0.2} \times 10^{42}$\,erg\,s$^{-1}$. All uncertainties are $1\sigma$.
\section{Discussion}
\subsection{Physical properties of the background galaxy}
\label{sec:properties}

The source is detected only in F160W, while the remaining four bands provide upper limits (Table~\ref{tab:photometry}). We use the delensed photometry at the geometric redshift, $z=2.14$, to constrain the stellar mass and star formation rate (SFR). At this redshift, the five HST bands sample rest-frame wavelengths of 1450\,\AA\ (F450W), 1875\,\AA\ (F606W), 2550\,\AA\ (F814W), 3360\,\AA\ (F105W), and 4895\,\AA\ (F160W). The three bluest bands therefore trace rest-UV emission from young massive stars, F105W lies just blueward of the Balmer and 4000\,\AA\ breaks, and F160W samples the rest-frame optical continuum, which traces the bulk of the stellar mass.

We converted the delensed flux densities to rest-frame luminosity densities as:
\begin{equation}
L_\nu(\lambda_{\rm rest}) = \frac{4\pi d_L^2 f_\nu(\lambda_{\rm obs})}{1+z},
\end{equation}
which we use for both the stellar mass (Sect.~\ref{sec:mass}) and the UV SFR limits (Sect.~\ref{sec:sfr}). For F160W, this gives $L_\nu(4895\,\mbox{\AA}) = 2.0 \pm 0.4 \times 10^{27}$\,erg\,s$^{-1}$\,Hz$^{-1}$, or $M_{\rm AB}(4895\,\mbox{\AA}) = -16.7 \pm 0.2$, corresponding to a rest-frame optical luminosity of $L_V \approx 3.9 \pm 0.8 \times 10^{8}\,L_\odot$.
\subsubsection{Stellar mass}
\label{sec:mass}

We estimated the stellar mass as $M_\ast = (M/L)_V\,L_V$, where the mass-to-light ratio $(M/L)_V$ depends on the age of the stellar population and on dust attenuation. Stellar population synthesis models \citep[e.g.,][]{2003MNRAS.344.1000B} give $(M/L)_V \approx 0.05$--0.1 for young ($\lesssim 100$\,Myr), dust-free populations, rising to $\approx 1$--2 for populations with ages of $\gtrsim 1$\,Gyr, the maximum possible at $z = 2.14$. The observed ${\rm F105W} - {\rm F160W}$ color is $> 1.3$\,mag at $3\sigma$, disfavoring a young, dust-free population that is expected to have an approximately flat spectrum across this range.

Two scenarios can reproduce the red color. In the first, it arises from the Balmer and 4000\,\AA\ breaks of an evolved, dust-poor population with $(M/L)_V \approx 1$--2, giving $\log(M_\ast/M_\odot) \approx 8.6$--8.9. In the second, the population is young but heavily obscured. For the \citet{2000ApJ...533..682C} attenuation curve, $E(3360 - 4895\,\mbox{\AA}) = 0.445\,A_V$, so the observed color requires $A_V \gtrsim 2.9$\,mag. Correcting the rest-frame 4895\,\AA\ luminosity for the corresponding attenuation raises the effective mass-to-light ratio of a young population from $\approx 0.1$ to $\gtrsim 2$, giving $\log(M_\ast/M_\odot) \gtrsim 8.9$. Since the resulting mass increases with $A_V$, we constrain it by using correlations between dust attenuation and stellar mass in star-forming galaxies at $z \sim 2$--3 \citep{2018MNRAS.476.3991M}. Galaxies with $\log(M_\ast/M_\odot) \sim 9$--9.5 typically have much lower attenuation than $A_V \approx 3$\,mag, and attenuation of $A_V \gtrsim 4$\,mag is characteristic of substantially more massive systems. Adopting $A_V \lesssim 4$\,mag as a physically reasonable upper limit gives $\log(M_\ast/M_\odot) \lesssim 9.4$ for the dusty scenario. Combining both scenarios, we adopt a stellar mass of $8.6 \lesssim \log(M_\ast/M_\odot) \lesssim 9.4$.

\subsubsection{Star formation rate}
\label{sec:sfr}

\begin{table}
\centering
\caption{Upper limits ($3\sigma$) on the unobscured star formation rate from the rest-frame UV non-detections, corrected for lensing magnification.}
\label{tab:sfr}
\begin{tabular}{lccc}
\hline
Filter & $\lambda_{\rm rest}$ & $\nu L_\nu$ & SFR$_{\rm UV}$ \\
 & (\AA) & (erg\,s$^{-1}$) & ($M_\odot$\,yr$^{-1}$) \\
\hline
F450W & 1450 & $< 1.9 \times 10^{41}$ & $< 0.009$ \\
F606W & 1875 & $< 3.5 \times 10^{41}$ & $< 0.016$ \\
F814W & 2550 & $< 3.5 \times 10^{41}$ & $< 0.024$ \\
\hline
\end{tabular}
\end{table}

The non-detections in the three rest-frame UV bands place upper limits on the unobscured SFR. We converted the $3\sigma$ limits on the rest-frame UV luminosity to SFR limits using the calibration constants in Table~1 of \citet{2012ARA&A..50..531K}:
\begin{equation}
\log\left(\frac{\rm SFR}{M_\odot\,{\rm yr}^{-1}}\right) = \log\left(\frac{\nu L_\nu}{{\rm erg\,s}^{-1}}\right) - \log C_x,
\end{equation}
with $\log C_x = 43.35$ for the far-UV (F450W and F606W) and $\log C_x = 43.17$ for the near-UV (F814W). The resulting $3\sigma$ limits are ${\rm SFR}_{\rm UV} < 0.009$, 0.016, and 0.024\,$M_\odot$\,yr$^{-1}$ from F450W, F606W, and F814W, respectively (Table~\ref{tab:sfr}).

For comparison, the star-forming main sequence of \citet{2014ApJS..214...15S} predicts ${\rm SFR} \approx 2$--10\,$M_\odot$\,yr$^{-1}$ at $z = 2.14$ for our stellar-mass range of $8.6 \lesssim \log(M_\ast/M_\odot) \lesssim 9.4$. The unobscured UV emission of our source is more than two orders of magnitude below this value. Alternatively, the galaxy may be heavily obscured. For the \citet{2000ApJ...533..682C} attenuation curve, $E(3360 - 4895\,\mbox{\AA}) = 0.445\,A_V$, so the observed ${\rm F105W} - {\rm F160W} > 1.3$\,mag requires $A_V \gtrsim 2.9$\,mag for an intrinsically flat young population. At this minimum attenuation, the dust-corrected limits are ${\rm SFR} \lesssim 9.6$, 7.1, and 4.1\,$M_\odot$\,yr$^{-1}$ from F450W, F606W, and F814W, respectively, consistent with a dust-obscured main-sequence galaxy. At the maximum plausible attenuation of $A_V \approx 4$\,mag, the limit rises to ${\rm SFR} \lesssim 30\,M_\odot$\,yr$^{-1}$.

\subsubsection{SED fitting}As an independent check, we fit the delensed photometry with
\texttt{Prospector} \citep{Johnson2021}, assuming a delayed-$\tau$ star
formation history at fixed $z = 2.14$ and including nebular emission. With a
single detection the posteriors are broad, giving
$\log(M_*/M_\odot) = 8.23^{+0.45}_{-1.35}$, $A_V = 0.79^{+0.39}_{-0.45}$\,mag,
and ${\rm SFR} = 1.5^{+17.4}_{-0.5}\,M_\odot\,{\rm yr}^{-1}$. These values
are broadly consistent with the estimates above.

\subsection{Origin of the X-ray emission}
\label{sec:xray_origin}

The X-ray luminosity of the source cannot be explained by star formation. At $z \approx 2$, the integrated emission from high-mass X-ray binaries scales as $L_X/{\rm SFR} \approx 1$--$2 \times 10^{40}$\,erg\,s$^{-1}\,(M_\odot\,{\rm yr}^{-1})^{-1}$ in the rest-frame 2--10\,keV band \citep{2016ApJ...825....7L}. Producing the observed $L_X = 1.2^{+0.3}_{-0.2} \times 10^{42}$\,erg\,s$^{-1}$ through star formation would therefore require ${\rm SFR} \approx 60$--120\,$M_\odot$\,yr$^{-1}$, significantly above the main-sequence expectation, more than three orders of magnitude above the unobscured UV limit, 15--30 times above the dust-corrected limit, and at least 2--4 times above the maximum plausible attenuation. We therefore conclude that the X-ray emission is dominated by an AGN, making the source a low-luminosity AGN hosted by a dwarf galaxy with $8.6 \lesssim \log(M_\ast/M_\odot) \lesssim 9.4$ at $z \approx 2.14$. The flat X-ray spectrum and the moderate intrinsic column density suggested by the spectral fit are consistent with a moderately absorbed AGN, as expected if the host galaxy is dusty.
%_____________________________________________________________
%                     Onecolumn continued float (place early!)
%-------------------------------------------------------------

% %_____________________________________________________________
% %                      Another onecolumn float (place early!)
% %-------------------------------------------------------------
% \begin{figure*}[ht!]
%    \resizebox{\hsize}{!}
%             {\includegraphics {figure.eps}}
%             %{\includegraphics[bb=10 20 100 100,clip]{figure.eps}}
%       \caption{Another onecolumn figure*}
%          \label{wide}
% \end{figure*}
% \twocolumn
% \clearpage
% % %%%%%%%%%%%%%%%%%%%%%%%%%%%%%%%%%%%%%%%%%%%%%%%%%%%%%%%%%%%%%%

%
%_____________________________________________________________
%                                             A rotated figure
%-------------------------------------------------------------

%%%%%%%%%%%%%%%%%%%%%%%%%%%%%%%%%%%%%%%%%%%%%%%%%%%%%%%%%%%%%%
\section{Conclusions}
We have presented the identification and multiwavelength characterization of a strongly lensed low-luminosity AGN at cosmic noon. By combining archival HST and Chandra observations with the strong-lensing models of SDSS J2243$-$0935, we constrain the source redshift, gravitational magnification, host-galaxy properties, and origin of the X-ray emission. Our main results are as follows.

\begin{enumerate}
\item We identify X-ray emission spatially coincident with all three HST images of the background source. Using the lens equation and the published lens-model realizations, we derive a geometric source redshift of $z=2.14^{+0.12}_{-0.17}$. This value is consistent with the interpretation of the system as a source at cosmic noon and provides an independent constraint relative to the previously reported photometric redshift.

\item The source is strongly magnified, with individual magnifications of $\mu_1=32.3^{+5.6}_{-4.5}$, $\mu_2=22.2^{+3.5}_{-2.6}$, and $\mu_3=22.0^{+6.0}_{-3.5}$. The consistency of the magnification-corrected F160W fluxes provides an independent check on the lensing reconstruction.

\item The intrinsic optical emission is extremely faint. The source is detected only in F160W, with $m_{\rm AB}=28.2\pm0.2$, while the other four bands provide upper limits. The resulting red F105W$-$F160W color can be produced either by an evolved stellar population or by substantial dust attenuation. We therefore infer a broad stellar-mass range of $8.6\lesssim\log(M_\star/M_\odot)\lesssim9.4$, with the lower end corresponding approximately to an evolved, relatively unobscured population and the upper end to a young, highly obscured population.

\item The non-detections in the rest-frame UV imply an unobscured star formation rate of $\mathrm{SFR}_{\rm UV}<0.009\,M_\odot\,\mathrm{yr}^{-1}$. If the dust is present, for the minimum attenuation required by the observed optical color, the dust-corrected SFR can reach several $M_\odot\,\mathrm{yr}^{-1}$, while larger attenuation permits substantially higher values.

\item The three lensed images contain $100\pm11$ net X-ray counts. The combined spectrum is consistent with a moderately absorbed power law, with $\Gamma=1.48^{+0.28}_{-0.25}$ and $N_{\rm H}=(5.5^{+5.2}_{-3.6})\times10^{22}\,\mathrm{cm}^{-2}$. Correcting for the total gravitational magnification gives an intrinsic rest-frame $2$--$10$ keV luminosity of $L_{\rm X}=1.2^{+0.3}_{-0.2}\times10^{42}\,\mathrm{erg\,s^{-1}}$.

\item The X-ray luminosity cannot arise from star formation-related processes alone, since it requires SFRs substantially above the derived limits. We therefore identify the X-ray source as an AGN and classify the system as a low-luminosity AGN hosted by a low-mass galaxy at $z\simeq2.14$.

\item This system illustrates the ability of strong gravitational lensing to expose intrinsically faint black-hole accretion at high redshift. The total magnification implies an effective Chandra exposure of approximately 9~Ms, deeper than the deepest Chandra fields. Additionally, the system represents an excellent target for in-depth follow-up studies to further constrain properties of LLAGN at the cosmic noon.

\end{enumerate}

%%%%%%%%%%%%%%%%%%%%%%%%%%%%%%%%%%%%%%%%%%%%%%%%%%%%%%%%%%%%%%
\begin{acknowledgements}
      MM thanks Keren Sharon for helpful insights.
\end{acknowledgements}

% Please note that we have included the references below in
% order to compile the document, but we ask you to:

\bibliographystyle{bibtex/aa}
\bibliography{bibtex/references}

\end{document}